\documentstyle[aps,preprint]{revtex}

\begin{document}
\draft
\title{ Tilting instability and other anomalies in the 
 flux-lattice in some magnetic superconductors
}
\author{T. K. Ng}
\address{Dept. of Physics, HKUST, Kowloon, Hong Kong}
\author{C.M. Varma}
\address{ Bell Laboratories, Lucent Technologies, Murray Hill, NJ07974}
\date{ \today }
\maketitle
\begin{abstract}
 The flux-line lattice in the compound
$ErNi_2B_2C$, which has a tendency to ferromagnetic order in the a-b plane
is studied with external magnetic field direction close to the c-axis. 
We show the existence of an instability where the direction of
flux-lines spontaneously tilts away from that of the applied field near
the onset of ferromagnetic order. The enhanced fluctuations in the flux 
lattice and the square flux lattice recently observed are explained 
and further experiments suggested. 
 \end{abstract} 


\narrowtext

   Recently, it was discovered that the flux-line lattice of the anisotropic 
magnetic superconductor $ErNi_2B_2C$ has very unusual properties.
In small angle neutron scattering experiments it was observed
that the vortex lattice structure (with magnetic
field applied at c-direction) is a square lattice as opposed to 
hexagonal as is typically found in a high field, type II
superconductor\cite{gam}. Moreover, for applied field $\vec{H}$
at an angle $\theta_a\sim1.6^o$ away from the c-axis, the total
field $\vec{B}$ is rotated towards the a-b plane and the
angle $\Delta\theta$ between $\vec{H}$ and total magnetic 
field $\vec{B}$ increases rapidly at low temperature\cite{gam}.
The FWHM of the rocking curve $\sigma_m$ which measures the
longitudinal correlation length $\xi_L$ of vortex lines
along their length is also found to increase sharply at low
temperature, with qualitatively similar temperature dependence
as the angle $\Delta\theta$\cite{gam}. 

   It is also observed that $ErNi_2B_2C$ has a
tendency of developing weak ferromagnetic order at low
temperatures $T\leq2.3K$\cite{e1,e3}. Such a 
transition cannot occur with a uniform  superconductive state 
preserved\cite{nv,t2,t1,t3,t4}.In a previous paper \cite{t2} we have 
suggested that a spontaneous vortex phase occurs in this material in the ferromagnetic state. The
magnetic properties of the material
can be described by the Ginsburg-Landau free 
energy functional\cite{t2} 
\begin{eqnarray}
\label{1}
{\em F} & = & \int{d^3r}[{1\over2}a|\psi|^2+{1\over4}b|\psi|^4+{\hbar^2\over
2m}|(\nabla-i{2e\over\hbar{c}}\vec{A})\psi|^2 + {\vec{B}^2\over8\pi}
\nonumber \\
 &  & +  
{1\over2}\alpha|\vec{M}|^2+{1\over4}\beta|\vec{M}|^4+{1\over2}\gamma^2
|\nabla\vec{M}|^2-\vec{B}.\vec{M}],
\end{eqnarray}
where $\vec{B}=\nabla\times\vec{A}$, $\vec{M}$ is magnetization and $\psi$
is the superconducting order parameter. The magnetic component $\vec{M}$ 
is found to be strongly anisotropic in $ErNi_2B_2C$, with magnetization
$\vec{M}$ reside essentially only along the in plane easy axis in
(100) and (010) directions\cite{e1}. We have shown 
that the unusual in-plane magnetic response of the compound
can be explained using the GL functional (1)\cite{nv}.
In this paper we shall study the out-of-plane
magnetic responses of $ErNi_2B_2C$ using the same GL functional (1).
We shall assume that the magnetization $\vec{M}$ lies
only on a-b plane and shall consider applied magnetic field making
small angle $\theta_a$ with the c-axis. The in-plane
anisotropy of the magnetic component is not included in the GL 
functional (1) but shall be considered later to understand
the square vortex lattice structure.

The competition between magnetism and superconductivity appears in
Eq. (1) as a Meissner effect of the superconducting
component towards the internal magnetic field produced by the
magnetic component $\vec{B}=4\pi\vec{M}$. For systems with
superconducting transition temperature $T_c$ higher than the magnetic
transition temperature $T_m$, the magnetic transition is suppressed.
However, the system may go through a second order phase transition
to a spiral phase or a first order transition to spontaneous
vortex phase at a slightly lower temperature $T_s<T_m$
\cite{t2,t1,t3,t4}. We shall concentrate at the temperature region
$T>T_{s(m)}$ in this paper and shall study changes in magnetic response
of the system as $T\rightarrow{T}_{s(m)}$. In this temperature
range $M$ is small and we can neglect the $M^4$ term in the GL
functional. The qualitative behaviour of the system at this
temperature range can be most easily understood by considering 
the London limit where $\psi=$ constant and neglecting the
$|\nabla\vec{M}|^2$ term in {\em F}\cite{nv}.

 Writing $\vec{B}=\vec{B}_c+\vec{B}_{ab}$, where $\vec{B}_c$ and
$\vec{B}_{ab}$ are the magnetic field along c direction and on
a-b plane, respectively, we obtain after minimizing {\em F} with 
respect to $\vec{M}$ and $\vec{A}$,
$\vec{M}=\vec{B}_{ab}/\alpha$, and $\vec{A}=\lambda_0^2
\nabla\times\vec{B'}$, where $\vec{B'}_z=\vec{B}_z$, and
$\vec{B'}_{ab}=(1-4\pi/\alpha)\vec{B}_{ab}$. Putting 
$\vec{M}$ and $\vec{A}$ back into $F$, we obtain
\begin{equation}
{\em F}\sim\int{d}^3r\left[{-a^2\over2b}+{1\over8\pi}
\left(\vec{B}_z^2+(1-{4\pi\over\alpha})\vec{B}_{ab}^2+\lambda_0^2(
\nabla\times\vec{B'})^2\right)\right],
\label{2}
\end{equation}
where $\lambda_0^2=mc^2/8\pi{e}^2|\psi|^2$ is the London penetration
depth for the 'pure' superconducting component. $\alpha$ is a decreasing
function of temperature and the magnetic transition (in the absence of
superconducting component) occurs at $\alpha(T_m)=4\pi$. Notice that
for magnetic field in the a-b plane, the presence of a
magnetic component reduces the overall cost in magnetic 
energy of the `pure` superconductor by a factor $(1-4\pi/\alpha)$ 
and also reduces the London penetration depth from
from $\lambda_0$ to $\lambda=(\sqrt{1-4\pi/\alpha})\lambda_0$\cite{nv}.
Notice that Eq.(2) can also be mapped into the problem of anisotropic
superconductors by rescaling variables. However, in the presence
of external field $H$, the two problems become very different as
we shall see later (see also ref.\cite{fv} and \cite{bl}).
To study the magnetic response, we first consider a single
vortex line solution in the London limit using Eq. (2). 
We shall assume that the vortex line is located in the a-c plane and makes a
small angle $\theta_v$ with the c-axis. Minimizing the free energy, we obtain
\begin{equation}
\label{3}
\lambda_0^2(\nabla\times\nabla\times\vec{B'})+\vec{B}=\hat{n}\Phi_0\delta
(y)\delta(x-ztan\theta_v),
\end{equation}
where $n_x=sin\theta_v, n_y=0$ and $n_z=cos\theta_v$. $\Phi_0$ is
the magnetic flux quantum. The equation can be further simplified
by using the vector identity $\nabla\times\nabla\times\vec{B'}=
\nabla(\nabla.\vec{B'})-\nabla^2\vec{B'}$ and the Maxwell equation
$\nabla.\vec{B}=0$. For small angle $\theta_v$, after some
algebra and transforming to momentum space, we obtain to order
O($\theta_v^2$), 
\begin{eqnarray}
\label{4}
B_z(\vec{q}) & = & {(1-\theta_v^2/2)\Phi_0\delta(q_z+\theta_vq_x)
\over1+\lambda_0^2
q^2+\theta_v^2\lambda^2q_x^2} \nonumber  \\
B_x(\vec{q}) & = & {\theta_v\Phi_0\delta(q_z+\theta_vq_x)
\over1+\lambda^2q^2}-
{4\pi\lambda_0^2\over\alpha}{\theta_vq_x^2\Phi_0\delta(q_z+\theta_vq_x)
\over(1+\lambda^2q^2)(1+\lambda_0^2q^2)}  \nonumber  \\
B_y(\vec{q}) & = & -{4\pi\lambda_0^2\over\alpha}{\theta_vq_xq_y\Phi_0
\delta(q_z+\theta_vq_x)\over(1+\lambda^2q^2)(1+\lambda_0^2q^2)}.
\end{eqnarray}
 
  The various terms in Eq. (4) can be understood as
follows: in the absence of the magnetic component ($\alpha
\rightarrow\infty$), only $B_z(\vec{q})$ and the first term in
$B_x(\vec{q})$ are non-zero and represents the magnetic field
of a vortex line tilted away from the c-axis on a-c plane with
small angle $\theta_v$. In
the presence of magnetic component on a-b plane, the 
magnetic response becomes anisotropic, leading to difference
in penetration depth in c and a-b directions ($\lambda_0$ and
$\lambda$, respectively) and 
distortion of vortex core where a small net magnetization in
a-direction is induced. As a result, a small magnetic dipolar
field is induced in the a-b plane which is represented by 
$B_y$ and the second term of $B_x$. The energy of single 
vortex line $\epsilon_1$ can be computed using
Eq. (2) and (4). In the limit
$(1-4\pi/\alpha)<<1$, we obtain to order $\theta_v^2$,
\begin{mathletters}
\begin{equation}
\label{5}
\epsilon_1=\epsilon_0(1+a_1\theta_v^2),
\end{equation}
where
\begin{equation}
\label{5}
a_1\sim{\pi\over\alpha}
({\lambda^2\over\xi^2}-2)(ln{\lambda_0\over\xi})^{-1}-
{4\pi\lambda^2\over\alpha\lambda_0^2},
\end{equation}
\end{mathletters}
where $\xi$ is the superconductor coherence length and $\epsilon_0\sim
(\Phi_0^2/4\pi\lambda_0^2)ln(\lambda_0/\xi)$ is the vortex line
energy when $\theta_v=0$. The first term in $a_1$ comes
from the induced magnetic moment in the vortex core
and from the corresponding dipolar field. This
term is positive ; 
the other correction terms are
negative and represent the lowering in energy from the magnetic
component when magnetic field is in the a-b plane\cite{nv}.
Similar analysis as can also be made when
the $|\nabla\vec{M}|^2$ term is included in the GL functional. We
find that the qualitative behaviour of the vortex solution is not modified
except that the divergence in $\lambda^{-1}$ as $T\rightarrow{T}_m$ 
is removed once the
$|\nabla\vec{M}|^2$ term is included. In particular, the
London penetration depth  saturates at a value of order $\lambda\sim
(\lambda_0\xi_m)^{1/2}$ as $T\rightarrow{T}_s$ for transition
to spiral state\cite{nv}, where $\xi_m\sim\gamma^2/\alpha$ is
the coherence length of the magnetic component.

  In the limit $H\sim{H}_{c1}$ where density of vortices is
low and interaction between vortices can be neglected, we may study
$\theta_v$ as function of angle of applied field to c-axis
$\theta_a$ using Eq. (5).
Consider the Gibb's energy functional 
\[
{\em G}={\em F}-\int{d}^3r{\vec{B}.\vec{H}\over4\pi}, \]
where $\vec{H}$ is applied field and the total magnetic 
field $\vec{B}$ is obtained by minimizing ${\em G}$ with
respect to $\vec{B}$.  For small angle $\theta_a$ and $\theta_v$,
Gibb's energy per unit volume is
\begin{equation}
\label{6}
{{\em G}\over{V}}\sim{B\over\Phi_0}\epsilon_0(1+a_1\theta_v^2)-{BH\over4\pi}
(1-{(\theta_v-\theta_a)^2\over2}).
\end{equation}
  Minimizing {\em G} with respect to $\theta_v$, we obtain
\[
\theta_v={\theta_a\over(1+2a_1(H_{c1}/H))}, \]
and
\[
H_{c1}(\theta_a)=H_{c1}(1+{a_1\theta_a^2\over(1+2a_1)})+O(\theta_a^4),
\]
where $H_{c1}=
4\pi\epsilon_0/\Phi_0$ is the lower critical field when the external
field is along c-axis ($\theta_a=0$). Notice that $\theta_v>\theta_a$ 
when $a_1<0$. This may occur when $\lambda/\xi$ is small enough, i.e. 
when the superconductor is not too strongly type II and when the  
system is close to magnetic instablity point $T\rightarrow{T}_{s}$.

   The above analysis can be easily extended to the intermediate
density regime $H\sim$ several $H_{c1}$ where the
density of vortices is of order $(2\pi\lambda_0^2)^{-1}$ and
the magnetic field $\vec{B}$ is already more or less uniform
in the superconductor. This is the regime of experimental
interests\cite{gam}. The magnetic field $\vec{B}$ in this case
can calculated using Eq. (3), except that the
right hand side of the equation is replaced by
$\hat{n}\Phi_0\sum_{\vec{R_n}}\delta(y-Y_n)\delta(x-X_n
-ztan\theta_v)$, where $\vec{R_n}$ are the positions of the vortices
in the vortex lattice. Following procedures similar to above (see
also ref\cite{tin} for details), we obtain in the small $\theta_a$
limit,
\begin{equation}
\label{7}
{{\em G}\over{V}}\sim{B^2\over8\pi}(1-{4\pi\over\alpha}\theta_v^2)
+{BH_{c1}\over4\pi}\left({ln(H_{c2}/B)\over{ln}(\lambda_0/\xi)}\right)
\left[1+a_1\theta_v^2\right]-
{BH\over4\pi}(1-{(\theta_v-\theta_a)^2\over2}),
\end{equation}
where $H_{c2}\sim\Phi_0/(2\pi\xi^2)$ is the upper critical field.
Neglecting Logarithmic corrections, we obtain after minimizing
{\em G} with respect to $B$ and $\theta_v$, $B\sim{H}-H_{c1}+B_o
\theta_v^2$ and
\begin{mathletters}
\begin{equation}
\label{8a}
\theta_v\sim{\theta_a\over{(1-{4\pi\over\alpha})+{H_{c1}\over{H}}(
{4\pi\over\alpha}+2a_1)}}.  
\end{equation}
where $B_o\sim(4\pi/\alpha)(1-2\pi/\alpha)H$ for $H>>H_{c1}$ and
the angle-dependent magnetization $M(\theta_a)$ is
\begin{equation}
\label{8b}
M(\theta_a)={B-H\over4\pi}\sim{B_o\theta_v^2-H_{c1}\over4\pi}.
\end{equation}
\end{mathletters}
Notice how $\theta_v$ is enhanced by the magnetic component in this
case. In particular, $\theta_v$ is always larger than
$\theta_a$ for $H>>H_{c1}$ and is diverging at low temperature
$T\leq{T}_m$, in contrast to the low density limit where
$\theta_v$ may be small than $\theta_a$ and is larger than
$\theta_a$ only when sufficiently close to magnetic
instability. Physically, the magnetic instability indicates
a magnetic transition where spontaneous magnetization along
a- or b- directions appears in the system at low
enough temperature, leading to a magnetic-field 
assisted spontaneous vortex phase\cite{nv} where the vortex
lattice tilts spontaneously from the c-axis below 
critial temperature  
\[
T_{sv}\sim{T}_m-{4\pi\over\alpha'}{H_{c1}\over{H}}(1+2a_1), \]
where $\alpha'=d\alpha/dT$. Direct observation of this
transition using neutron scattering or magnetic imaging
techniqiues is suggested.

  Next we consider the case $\theta_a=0$ and study thermal 
fluctuations in positions of vortex lines along the c-direction
in the intermediate density regime. We
consider the model free energy {\em $F_N$} for $N$
vortex lines in a sample of thickness $L$, defined by their
trajectories $[\vec{r}_j(z)]$ as they traverse through sample with
external magnetic field along the z-axis,
\begin{eqnarray}
\label{9}
{\em F_N} & = & {1\over2}\epsilon_{t1}\sum_{j=1}^N\int_0^L\left({
d\vec{r}_j(z)\over{d}z}\right)^2dz+{1\over2}\sum_{i\neq{j}}
\int_0^L\epsilon_t(|\vec{r}_i(z)-\vec{r}_j(z)|)
\left({d\vec{r}_i(z)\over{d}z}-{d\vec{r}_j(z)\over{d}z}
\right)^2dz  \nonumber  \\
  &  &  + \sum_{j=1}^N\int_0^LV_D[\vec{r}_j(z)]dz+
{1\over2}\sum_{i\neq{j}}\int_0^L
V(|\vec{r}_i(z)-\vec{r}_j(z)|)dz
\end{eqnarray}
where $V_D(\vec{r}(z))$ is a random pinning potential for vortex lines
and $V(r)\sim\epsilon_0K_0(r/\lambda_0)$ is the interaction potential
between vortex lines. In the following we shall replace $V_D(\vec{r})$
by the {\em average} pinning potential $V_D(\vec{r}_j(z))\sim
\kappa_0(\vec{r}_j-\vec{R}_j)^2/2$ where $\vec{R}_j$ are sites on
the flux lattice. We shall also expand the interaction potential
around it's minimum point $V(r)\sim\epsilon_0(r/d)^2$ where
$d\sim(\Phi_0/B)^{1/2}$ is the distance between vortex lattice
sites. Note that $d\vec{r}_j(z)/dz\sim\theta_v(\vec{r}_j)$ in 
our analysis and the first two terms in
Eq. (9) come from terms proportional to
$\theta_v^2$ in the Gibb's energy (7).
The tilt modulus $\epsilon_{t1}$ is chosen such that when
all $\theta_v(\vec{r}_j)$'s are equal, the energy is given
by Eq. (7) $\epsilon_t(r)$ can be extracted from
the free energy {\em G} of a flux-line lattice configuration 
where all but one of the flix-lines have angle
$\theta_v(\vec{r_j})=0$. The field configuration in this case can be 
calculated
from an appropriate combination of eqn.(4) and the field configuration
leading to eqn.(7) with $\theta=0$. We shall assume
that the flux lines form a square lattice and retain
only the nearst neighbor terms in the sum $i\neq{j}$ in computing
$\epsilon_t(r)$. The qualitative properties of $F_{N}$ do not
depend sensitively on this approximation. We also ignore the ordinary contribution to the tilt moduli expressing the effect of increased line-length
as it is negligible compared to the effects we are interested in. We obtain
\begin{mathletters}
\label{masses}
\begin{equation}
\label{101}
\epsilon_{t1}\sim\epsilon_0({4\pi\over\alpha}+2a_1+
(1-{4\pi\over\alpha}){H\over{H}_{c1}}),
\end{equation}
where we have neglected the logarithmic correction terms and 
$\epsilon_t(r)=\epsilon_{t2}\delta(r-d)$, where
\begin{equation}
\label{102}
\epsilon_{t2}\sim{\Phi_0^2\over4\alpha\lambda_0^2}(ln{H_{c2}\over{B}})
({H\over{H}_{c1}}-1).
\end{equation}
\end{mathletters}
Notice that $\theta_v/\theta_a=(H/H_{c1})(\epsilon_0/\epsilon_{t1})$.
Notice also that at low temperature $T\leq{T}_m, 
\epsilon_{t1}$ is a decreasing function of vortex density.
In particular, $\epsilon_{t1}\rightarrow0$ at the spontaneous
vortex phase transition temperature $T\rightarrow{T}_{sv}$.

The thermodynamic properties of $F_N$ can be obtained
rather easily by observing that the variable
z can be treated as an imaginary time and the model
can be mapped into a quantum mechanical problem of coupled
harmonic oscillators on a square lattice with
$\hbar\rightarrow{T}$\cite{nel}.  The Hamiltonian of
our effective quantum mechanical system is in fourier space,
\begin{equation}
\label{11}
H=\sum_{\mu,\vec{k}}{1\over2m(\vec{k})}P_{\mu}(\vec{k})P_{\mu}(-\vec{k})+
\sum_{\mu,\vec{k}}{m(\vec{k})\omega(\vec{k})^2\over2}
X_{\mu}(\vec{k})X_{\mu}(-\vec{k}),
\end{equation}
where $X_{\mu}(\vec{k})=N^{-1/2}\sum_{i}e^{-i\vec{k}.\vec{R_i}}r_{i\mu}$,
$\mu=\hat{x},\hat{y}$ and $P_{\mu}(\vec{k})=
-iT\partial/\partial{X}_{\mu}(\vec{k})$ is the canonical momentum 
conjugate to $X_{\mu}(\vec{k})$. The momentum dependent mass 
$m(\vec{k})$ and frequency $\omega(\vec{k})$ are
\[
m(\vec{k})=\epsilon_{t1}+4\epsilon_{t2}\gamma(\vec{k}), \]
and
\[
\omega(\vec{k})=\sqrt{{\kappa_0+4{\epsilon_0\over{d}^2}\gamma(\vec{k})
\over{m}(\vec{k})}},  \]
where $\gamma(\vec{k})=sin^2(k_xd/2)+sin^2(k_yd/2)$. The thermodynamic
properties of the flux-line lattice can be obtained easily from (11).
We obtain
\[
<X_{\mu}(\vec{k},z)X_{\nu}(\vec{k},z')>=\delta_{\mu\nu}{T\over
2m(\vec{k})\omega(\vec{k})}e^{-\omega(\vec{k})|z-z'|},  \]
and
\begin{equation}
\label{12}
<r^2>={1\over{N}}\sum_{\mu,\vec{k}}{T\over2m(\vec{k})\omega(\vec{k})}
\sim{Td\over\sqrt{(\kappa_0d^2)\epsilon_{t2}+\epsilon_0\epsilon_{t1}}}
\end{equation}
where $<r^2>$ is the mean square displace of flux lines from their
equilibrium positions, $\kappa_0d^2\sim$ average pinning energy of
the vortex lines per unit length. Notice that $<r^2>$ remains finite
when $\epsilon_{t1}\rightarrow0$ as long as there is a finite
pinning strength for the flux-line lattice.  The longitudinal
correlation length measured in neutron scattering experiment is $\xi_L\sim
\omega(\vec{k}\rightarrow0)^{-1}\sim\sqrt{\epsilon_{t1}/\kappa_0}$, and
$\sigma_m\sim{d}/\xi_L$, which diverges as $(\epsilon_{t1})^{-1/2}$ as 
the instability towards spontaneous vortex phase is approached. In
particular, we obtain a scaling relation between $\sigma_m$ and
$\theta_v/\theta_a$,
\begin{equation}
\label{13}
\sigma_m^2\left({\theta_a\over\theta_v}\right)\sim({H_{c1}\over{H}})
({\kappa_0d^2\over\epsilon_0}). 
\end{equation}
While Eq. (8) and (13) are in good agreement with the existing data
in Ref. (1), data over a range of fields and lower temperatures is
required to test the theory. Notice that our theory for thermal fluctuations
is essentially a Gaussian theory and Eq.(8) and (13) are essentially
mean-field results. Derviation from mean-field behaviour
is expected at temperature very close to transition temperature
$T_{sv}$ when the system is in the critical regime. We have estimated
the size of the critical regime using the specific heat obtained
from our theory and find that the critical regime is given by
\[
t\leq\left({\kappa_od^2\over\epsilon_o}\right)^{1\over3}
\left({1\over{d}^3\Delta{C}}\right)^{2\over3}, \]
where $t=|1-T/T_{sv}|$ and $\Delta{C}$ is the specific heat jump
across the transition in mean-field theory. The size of the critical regime
is expected to be very small at intermediate density of vortices
$d\sim\lambda_0$.

  Finally we make a plausible suggestion on the origin of square
lattice structure of the flux-line lattice observed in 
neutron scattering experiment. Another possible origin is
Fermi surface anisotropy which leads to anisotropic shape of
vortex core. As is evident from our discussions,
the enhancement of fluctuations in positions of vortex lines
is a result of enhanced magnetic response arising from
the magnetic component in the system. In
$ErNi_2B_2C$ compound, the magnetic component is strongly
anisotropic in a-b plane, with magnetization $\vec{M}$ reside
essentially only along the a- or b- axes. Thus we expect that
in a treatment of the system with anisotropy effect included,
the fluctuations in vortex line positions, and
correspondingly, the mean-square displacement of vortex lines,
$<r_ir_j>$ will also be anisotropic and enhanced more in a- and b-
directions.  Effectively, the shape of the vortex core becomes
"cross-like", with vortex core elongated along a- and b-
directions. When density of vortices is high enough, the repulsion
between vortex cores becomes important in determining the vortex
lattice structure. In particular, because of the small
energy difference between square and triangular lattices\cite{tin},
it is reasonable that the vortices will arrange in
a square lattice with angle $\pi/4$ with respect to a- and b-
axes to maximize distances between each other. This structure
is indeed what is observed experimentally\cite{gam}.

  Notice however that the anisotropy in magnetic response is absent
in the linear response regime and comes in only through the $M^4$
term in the GL functional. Thus the anisotropy effect just discussed
 is expected
to be strong only at temperature regime
$T\leq{T}_m$, when the magnetic linear response is
close to diverging and the $M^4$ term becomes important. 
Experimentally, a square vortex lattice is observed at
temperature well above the instability temperature determined
by the divergence in $\theta_v/\theta_a$ or $\sigma_m$.  
To explain this we suggest the 
interesting possibility that the anisotropy
effect are strong even at $T-T_m$ comparable to or larger than
$T_m$ if the size of the
vortex core, $\xi$ is microscopically large (the size
of vortex core $\sim\xi$, is found to be $135\AA$\cite{gam} in 
$ErNi_2B_2C$ compound). In this case,
the behaviour of vortex core can be described by GL functional (1)
with $\psi\sim0$ and has an magnetic instability at
$T=T_m'$, where $T_m'\geq{T}_m$ when terms of form $\eta\vec{M}^2
|\psi|^2$ which describes effect of conduction-electron 
polarization on superconductivity are present in the
system\cite{t1}. In this case, spontaneous magnetization 
occurs inside vortex cores at temperature $T>T_m$, pointing
randomly in the  $\pm{a}$ and $\pm{b}$ directions\cite{1d}, leading to 
anisotropic fluctuation effect observed well above $T_m$. 

   The square lattice structure is stabilized only when the vortex-vortex interaction is sensitive to the core-shape. This can happen only for
vortex density above a critical value. Indeed, the square lattice is
found only above a critical field below which it has the conventional
hexagonal shape.

   This line of argument suggests also that the square flux-line lattice
will disappear and replaced by a regular triangular lattice if the
external field is applied along (010) or (100) directions. It also
predicts that the triangular lattice will become distorted if the
external field is applied on (110) direction. In the later case
the vortex core will be effectively elongated along a-b plane. 
Correspondingly, the distance between
vortices joint by a line in a-b plane will be larger than
distance between vortices joint otherwise. Experimental
observation of flux-line lattice with applied field  
in (100) or (110) directions are suggested.

   We wish to acknowledge very useful discussions with Peter Gammel,D.Huse
and U. Yaron. T.K. Ng acknowledge partial support by UGC Hong Kong,
through RGC Grant no.UST623/95P.


\begin{references}
\bibitem{1d} Notice that magnetization residing in core of a vortex-line
    forms an effective one-dimensional system which is
    always disordered at finite temperature $T>0$.
\end{references}
\end{document}